\RequirePackage{fix-cm}
\documentclass[twocolumn,epjc3]{svjour3}
\smartqed  
\RequirePackage{graphicx}

\usepackage{amsmath,amssymb,graphics}
\usepackage{epsfig,subfigure}
\usepackage{color}

\journalname{Eur. Phys. J. C}
\begin{document}

\title{Brane worlds in gravity with auxiliary fields}



\author{Bin Guo\thanksref{e1}
        \and Yu-Xiao Liu\thanksref{e2}
         \and Ke Yang\thanksref{e3}
        }

\thankstext{e1}{e-mail:guob12@lzu.edu.cn}
\thankstext{e2}{e-mail:liuyx@lzu.edu.cn, corresponding author}
\thankstext{e3}{e-mail:yangke09@lzu.edu.cn}


\institute{Institute of Theoretical Physics,
            Lanzhou University, Lanzhou 730000,
            People's Republic of China
}

\date{Received: date / Accepted: date}

\maketitle

\begin{abstract}
Recently, Pani, Sotiriou, and Vernieri explored a new theory of gravity by adding
nondynamical fields, i.e., gravity with auxiliary fields [Phys. Rev. D 88, 121502(R) (2013)]. In this gravity theory, higher-order derivatives of matter fields generically appear in the field equations. In this paper we extend this theory to any dimensions and discuss the thick braneworld model in five dimensions. Domain wall solutions are obtained numerically. The stability of the brane system under the tensor perturbation is analyzed. We find that the system is stable under the tensor perturbation and the gravity zero mode is localized on the brane. Therefore, the four-dimensional Newtonian potential can be realized on the brane.
\end{abstract}

\section{Introduction}\label{SS1}

In order to solve the problems such as singularity, nonrenormalizability, dark energy, and dark matter in general relativity, modified gravity theories have been presented and investigated.
For recent reviews of modified gravities and related topics, see Refs. \cite{Nojiri2011PhysRept,Clifton2012,Capozziello:2011et}. One of such modifications studied extensively is the Palatini extension of the modified gravity. For the Einstein-Hilbert action, the Palatini theory is the same as the original metric theory. But if the action differs from the Einstein-Hilbert action, one usually gets a different gravity theory. Palatini $f(R)$ \cite{Sotiriou2010}, Eddington-inspired Born-Infeld (EiBI)\cite{Banados2010}, and Born-Infeld-$f(R)$ \cite{Makarenko2014} gravities are known such theories.
Interestingly, it was found that the EiBI gravity is identical to a bimetric gravity with an auxiliary field  \cite{Delsate2012}.
These theories can be considered as adding auxiliary fields to the action, and attract much attention in recent years. Besides, adding auxiliary fields is always helpful to construct Lagrangian formalism of some theories.

Recently, Pani, Sotiriou, and Vernieri suggested a new gravity theory that modifies general relativity by adding nondynamical auxiliary fields \cite{Pani2013}. This theory satisfies the weak equivalence principle and the corresponding modified Einstein equations contain higher-order derivatives of the matter fields. Although the details of the nature of the auxiliary fields and the way they enter to the action are unclear, this work provides a generic framework to study the phenomena and observational constrains of the gravity theory with auxiliary fields.
This theory is determined by only two parameters up to the next to
leading order in the derivative expansion \cite{Pani2013}. And in some approximations, EiBI and
Palatini $f(R)$ gravities correspond to the special cases of the theory.
For gravity with tensor auxiliary fields,
see a subsequent note by Ba$\tilde{\text{n}}$ados and Cohen \cite{Banados:2013vya}.

On the other hand, the extra dimension theory gives a new view of our universe, and opens a new way to solve the gauge hierarchy and cosmology problems. One of the famous models of this theory is the Randall-Sundrum (RS) braneworld model \cite{Randall1999,Randall1999a}, which is considered in general relativity. It provides the theoretical predictions of extra dimension effects, which may be detected in future experiments and observations. Along with the progress in the modified gravity and the braneworld models, there are many efforts to put to the braneworld model in modified gravities \cite{Giovannini:2001ta,Arias2002,Barbosa-Cendejas2005,Barbosa-Cendejas2006,Liu2010a,Liu2011a,Liu2011,HoffdaSilva2011,Carames2012,
Antoniadis2012,Ahmed2012,Yang2012b,Bazeia2012,Liu2012,Liu2012a,Yang2012a,Guo2012,Liu2013,German:2013sk,Bazeia2013,Bazeia2014}.
In this paper, we will apply the gravity theory with auxiliary fields to braneworld model and investigate the deformation and stability problems of the brane system. We wish this work will shed light on future studies of applying certain gravity theory with auxiliary fields to braneworld models.

For a thin brane model, the energy-momentum tensor of the brane is a delta function of the extra dimension.
In the original RS model, the field equations are second order and hence the thickness of the brane can be neglected. However, if a gravity theory contains higher-order derivatives of matter fields, such as the gravity theory with auxiliary fields, it is very hard to solve the field equations for a thin brane model.
On the other hand, the thick brane model, which can be used to study the inner structure of the brane, is a nature extension of the RS model \cite{DeWolfe2000,Gremm2000a,Csaki2000}. The brane configuration is usually generated by a smooth scalar field which connects two nontrivial vacua \cite{Dzhunushaliev2010}. So the energy momentum tensor of the matter field is a smooth function of the extra dimension and it is convenient to apply to the gravity theory with auxiliary fields.

Theoretically, the important problems in a braneworld model include the stability problem and the localization of the massless graviton on the brane, which are essential to recover the effective four-dimensional Newtonian potential on the brane. Experimentally, the interest mainly focuses on the phenomenology of braneworld models, such as the deviation from the Newton potential caused by the massive Kaluza-Klein (KK) gravitons, and the high-energy particle scattering process involving KK particles.
However, the spectrum of the gravity KK modes is determined by the brane configuration, which depends on the braneworld model.
In this paper, we study the braneworld model in the new gravity theory with auxiliary fields.
We find that domain wall solution is supported in this theory.
The brane system is stable under the tensor perturbation, and the massless graviton is localized on the brane.
Furthermore, we also find some new phenomena that do not appear in general relativity.


\section{Brane model with auxiliary fields and numerical solutions}\label{model and background equations}

Following Ref. \cite{Pani2013}, we start with the gravity theory with auxiliary fields in the level of the equations of motion in $D$-dimensional spacetime,
\begin{eqnarray}\label{Field_Eq_1}
R_{AB}-\frac{1}{2}R g_{AB}=T_{AB}+S_{AB}[T_{CD},g_{CD}], \label{FieldEqs}
\end{eqnarray}
where the tensor $S_{AB}$ is constructed from the energy-momentum tensor $T_{AB}$. We use
$A,~B,~C,\cdots$ to denote the bulk index $0, 1, 2, \cdots,D-1$.
To keep the weak equivalence principle, we need $\nabla_{A}T^{AB}=0$.
Then the Bianchi identity implies $\nabla_{A}S^{AB}=0$. Besides, the tensor $S_{AB}$ should vanish when $T_{AB}=0$. As the matter field equations are unchanged, it is hard to construct the Lagrangian for the matter part.

Generally, the energy-momentum tensor $T_{AB}$ for scalar fields contains second derivatives, so the most general form of $S_{AB}$ up to fourth order in derivatives is \cite{Pani2013}
\begin{eqnarray}
S_{AB}&=&\alpha_1g_{AB}T  \nonumber \allowdisplaybreaks\\
&&+\alpha_2g_{AB}T^2+\alpha_3TT_{AB}+\alpha_4g_{AB}T_{CD}T^{CD}  \nonumber\allowdisplaybreaks \\
&&+\alpha_5T^{C}_{~A}T_{CB}+\beta_1\nabla_{A}\nabla_{B}T+\beta_2g_{AB}\Box T \nonumber\allowdisplaybreaks\\
&&+\beta_3\Box T_{AB}+2\beta_4\nabla^C\nabla_{(A}T_{B)C}+\cdots, \allowdisplaybreaks\label{S_AB}
\end{eqnarray}
where $T=g^{AB}T_{AB}$, and the possible terms containing the
Levi-Civita tensor are not considered because they would violate parity.
The requirement $\nabla_{A}S^{AB}=0$ imposes some relations between the coefficients $\alpha_i$ and $\beta_i$, which will be determined by the perturbation method. Using the following relations \cite{Pani2013}
\begin{eqnarray}
 \left(\Box\nabla_{B}-\nabla_{B}\Box\right)T &=& R_{AB}\nabla^{A}T,\\
 \left(\nabla^{A}\nabla^{C}\nabla_{A}-\nabla^{C}\Box\right)T_{CB}
   &=& R_{ABCD}\nabla^{D}T^{CA},\\
 \nabla^{A}R_{ABCD} &=& 2\nabla_{[C}R_{D]B},
\end{eqnarray}
and the lowest-order equations $R_{AB}-\frac{1}{2}R g_{AB}=T_{AB}+\alpha_1 T g_{AB}
$, the higher derivatives in $\nabla_{A}S^{AB}=0$ can be eliminated and the relations between the coefficients $\alpha_i$ and $\beta_i$ turn out to be
\begin{subequations}\label{coefficients}
\begin{eqnarray}
\alpha_1 \!\!&=&\!\! 0, ~
\alpha_2 =\frac{\beta_1-\beta_4}{2(D-2)}, ~
\alpha_3 =\frac{2\beta_4}{D-2}-\beta_1, \\
\alpha_4 \!\!&=&\!\! \frac{1}{2}\beta_4, ~
\alpha_5 =-2\beta_4, ~
\beta_2  =-\beta_1, ~
\beta_3=-\beta_4.
\end{eqnarray}
\end{subequations}
The above results are a generalization of Ref. [8] for $\Lambda=0$ in $D$-dimensional spacetime.
So there are only two free parameters $\beta_1$ and $\beta_4$.
Note that the field equations (\ref{FieldEqs})-(\ref{S_AB}) and the above relations are meaningful only within a derivative expansion which is truncated to some order because they are determined by the perturbation method.
Likewise, the tensor $S_{AB}$ is
divergenceless ($\nabla_{A}S^{AB}=0$) only to the same order in the derivative expansion, and
quantitative results can be derived to a given order in such expansion.


In this paper, we investigate the thick brane model in the above gravity theory with auxiliary fields in five-dimensional spacetime ($D=5$). The metric for a flat brane is assumed as
\begin{eqnarray}\label{metric}
ds^2=g_{AB}dx^Adx^B=a^2(z)(\eta_{\mu\nu}dx^{\mu}dx^{\nu}+dz^2),
\end{eqnarray}
where $\mu,\nu$ denote the brane coordinate indices $0,1,2,3$, $\eta_{\mu\nu}=\text{diag}(-,+,+,+)$, $z$ is the extra dimension coordinate, and $a^2(z)=e^{2A(z)}$ is the warp factor.

In our brane model, the brane is generated by a background scalar field $\phi$ with the Lagrangian density ${\cal{L}}_{\phi}=-\frac{1}{2}\partial_A\phi \partial^A\phi-V(\phi)$.
Hence, the energy-momentum tensor takes the form:
\begin{eqnarray}
T_{AB}=\partial_A\phi\partial_B\phi-\frac{1}{2}g_{AB}\partial_C\phi\partial^C\phi-g_{AB}V(\phi),
\end{eqnarray}
where the scalar field $\phi=\phi(z)$ is a function of $z$ for a static brane solution.
The equation of motion for the scalar field is given by
\begin{eqnarray}
V'=e^{-2A}(\phi''+3A'\phi')\phi', \label{ScalarEOM}
\end{eqnarray}
where the prime denotes the derivative respect to the extra dimension coordinate $z$.

The $\mu\mu$ and $55$ components of the Einstein equations take the form
\begin{eqnarray}
uV^{2}+fV+g&=&0, \label{implicit eq 1}\\
uV^{2}+hV+s&=&0, \label{implicit eq 2}
\end{eqnarray}
where $u$, $f$, $g$, $h$, and $s$ are expressions of $A$, $A'$, $A''$, $\phi$, $\phi'$, $\phi''$, and $\phi'''$:
\begin{eqnarray}
u&=& \frac{1}{6}(5 \beta_1 + 2 \beta_4 ) e^{2A} ,\allowdisplaybreaks\\
f&=&e^{2 A} \!+\! \frac{1}{6}(9 \beta_1 + 2 \beta_4 ) \phi'^2, \allowdisplaybreaks\\
g&=& 3(A'^{2}+A'')
   + \frac{1}{2}\phi'^2
   +\frac{1}{4}e^{-2A}
 \Big[-9 \beta_4 A'^2 \phi'^2         \nonumber \allowdisplaybreaks\\
  && +    \frac{1}{2}(3 \beta_1  - 2\beta_4) \phi'^4
    - (17 \beta_1 + 5 \beta_4 )(\phi''^2 + \phi'\phi''')
       \nonumber \allowdisplaybreaks\\
    &&-  (3 \beta_1 + 7 \beta_4 )A''\phi'^2
     -  (6\beta_1 + 9 \beta_4) A'\phi' \phi''
       \Big], \allowdisplaybreaks\\
h&=&e^{2A} \!-\! \frac{1}{6}(21\beta_1 + 2 \beta_4 ) \phi'^2, \allowdisplaybreaks\\
s&=& 6A'^{2}- \frac{1}{2} \phi'^2
 +  \frac{1}{4}e^{-2A}
\Big[ - 3 (4 \beta_1 - 5 \beta_4)A'^2  \phi'^2\nonumber \allowdisplaybreaks \\
&&- \frac{3}{2} (3 \beta_1 - 2 \beta_4 ) \phi'^4
    -3\beta_4 (\phi''^2+ \phi'\phi''')  \nonumber \allowdisplaybreaks\\
&&+ 7\beta_4 A''\phi'^2  -(68 \beta_1 -3 \beta_4) A'\phi' \phi''      \Big] . \allowdisplaybreaks
  \end{eqnarray}
In the above equations, we have used the exact expression of $V'$ in Eq.~(\ref{ScalarEOM}) and its derivative to eliminate the $V'$ and $V''$ terms.

Usually, if the Einstein equations are second order (the case of $S_{MN}=0$),
one can achieve a topological nontrivial domain wall solution by introducing a superpotential \cite{DeWolfe2000,Gremm2000a,Sasakura2002,Afonso2006}, where the scalar field has a kink configuration and it connects the two degenerate vacua of the self-interaction potential $V(\phi)$ \cite{Rubakov2001}, and the spacetime is asymptotically anti-de Sitter (AdS) at the boundary of the extra dimension.
But it is hard to solve the above equations (\ref{ScalarEOM})-(\ref{implicit eq 2}) for a given scalar potential $V(\phi)$ or a given warp factor $A(z)$ even numerically, because of the appearance of higher derivatives of $\phi$ and many nonlinear terms. However, we can also expect that the spacetime is also asymptotically anti-de Sitter (AdS) when $z\rightarrow\pm\infty$, for which the scalar field $\phi(z)\rightarrow{\pm}v$ and hence $T_{MN}$ and $S_{MN}$ vanish at the boundary. Therefore, for simplicity, we assume that the scalar field has the following kink configuration \cite{Kehagias2001,Liu2011,Guo2012}
\begin{eqnarray}
\phi(z)=v\tanh(k z) \label{phi}
\end{eqnarray}
and solve numerically the scalar potential and the warp factor.
We will show later that this simple choice leads to interesting results.
In order to guarantee the validity of the perturbation method,
the expansion coefficients $\beta_1$ and $\beta_4$ should be small enough.


\begin{figure}[htbp]
\includegraphics[scale=0.41]{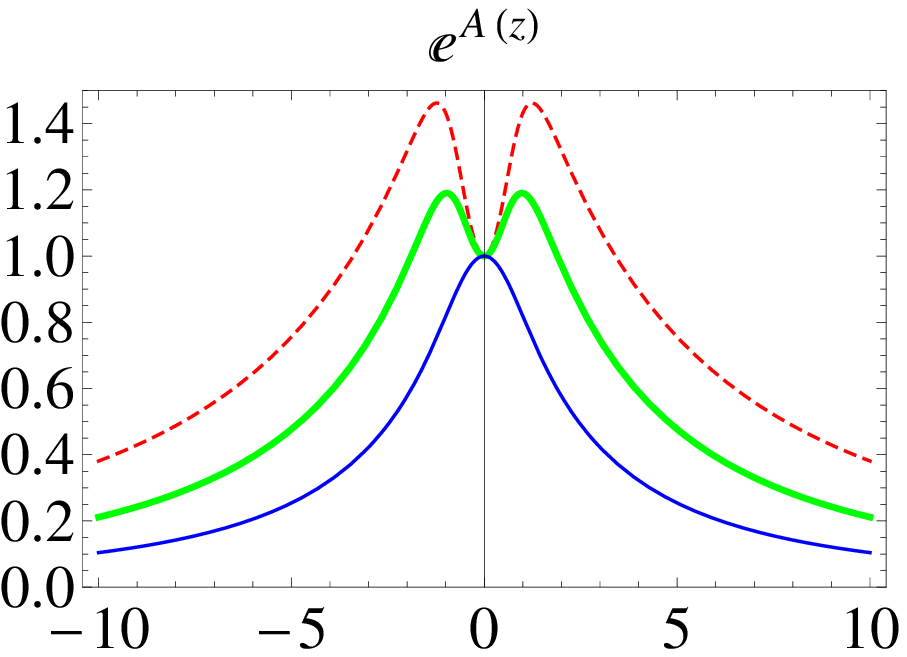}
\includegraphics[scale=0.45]{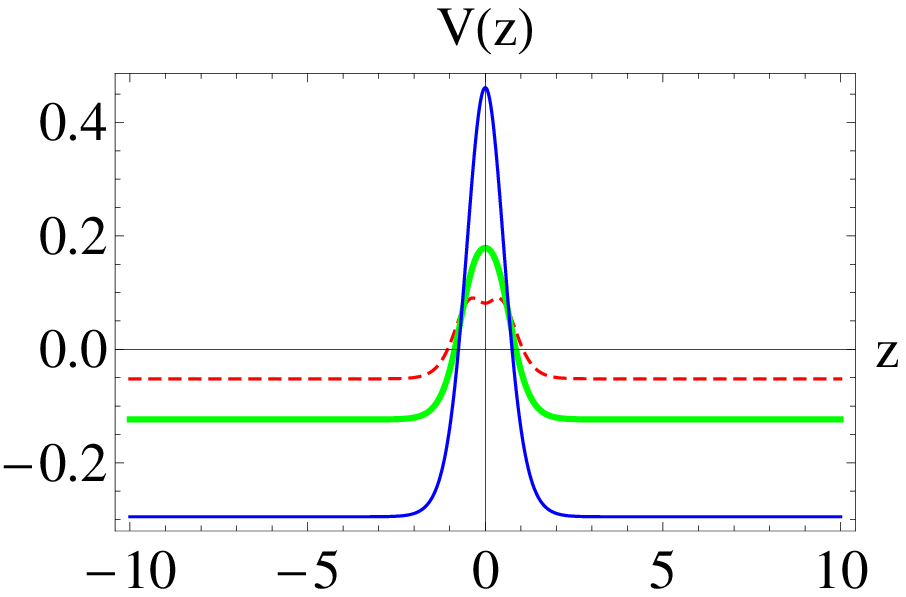}
\includegraphics[scale=0.45]{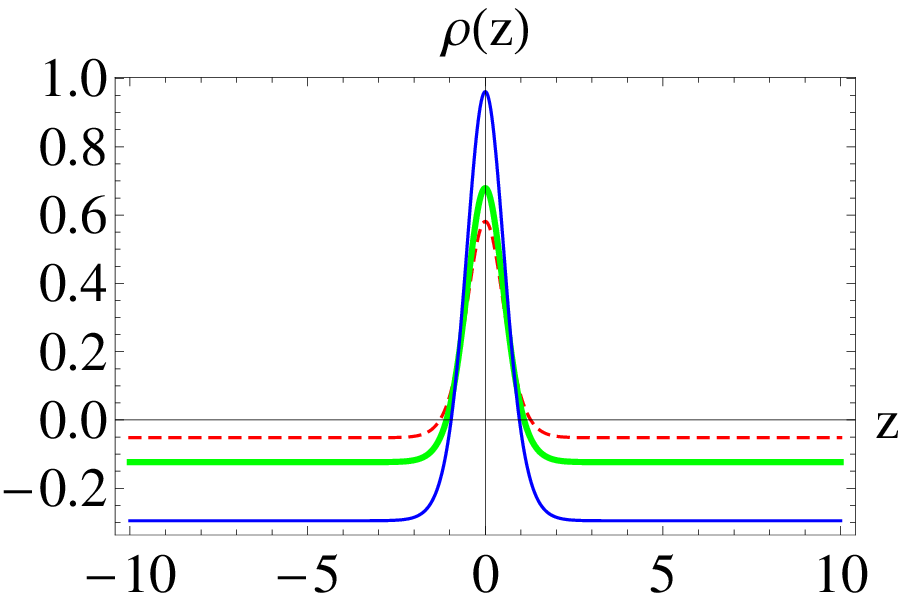}
\includegraphics[scale=0.45]{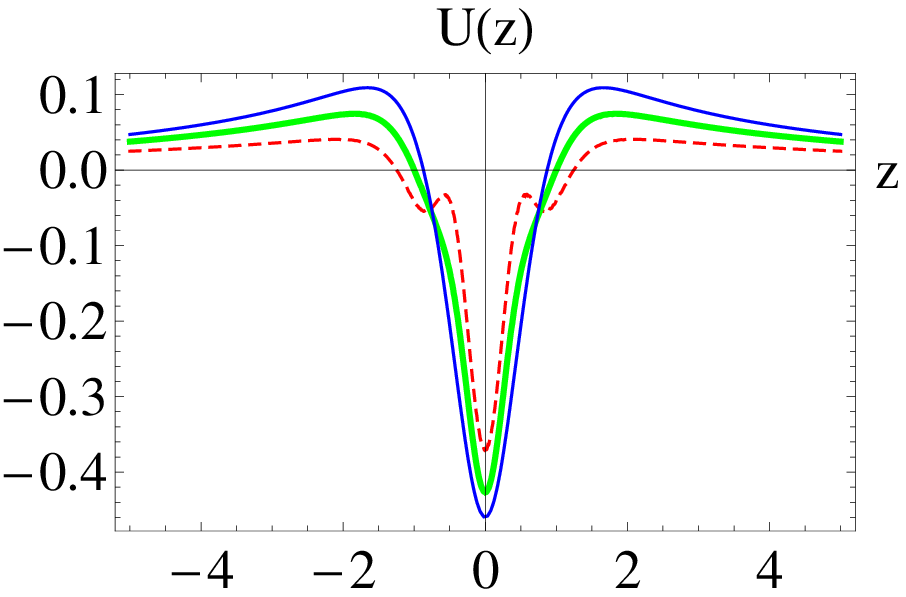}
\caption{The shapes of the warp factor $A(z)$, scalar potential $V(\phi(z))$, energy density $\rho(z)$, and effective potential of the gravity KK modes $U(z)$ corresponding to the solution $A_+(z)$. The parameters are set to $D=5$, $k=1$, $v=1$, $(\beta_1,\beta_4)=(-0.4,-0.1)$ (dashing red lines), $(-0.2,-0.1)$ (thick green lines) and $(0.01,-0.1)$ (thin blue lines). \label{picture 1} }
\end{figure}

With the assumption (\ref{phi}) for the scalar field, $u$, $f$, $g$, $h$, and $s$ are only functions of $A$, $A'$, and $A''$. Subtracting Eq.~(\ref{implicit eq 1}) from Eq.~(\ref{implicit eq 2}) yields the expression for the potential:
\begin{eqnarray}
V=\frac{s-g}{f-h}. \label{V}
\end{eqnarray}
Then, by substituting Eq.~(\ref{V}) into Eq.~(\ref{implicit eq 1}) or (\ref{implicit eq 2}),
we finally get a second-order ordinary differential equation for the warp factor $A(z)$:
\begin{eqnarray}
u(g - s)^2 +(f - h)( fs-gh)=0, \label{WarpFactorEOM}
\end{eqnarray}
which can be solved numerically by introducing the boundary conditions $A(0)=A'(0)=0$.

\begin{figure}[htbp]
\includegraphics[scale=0.41]{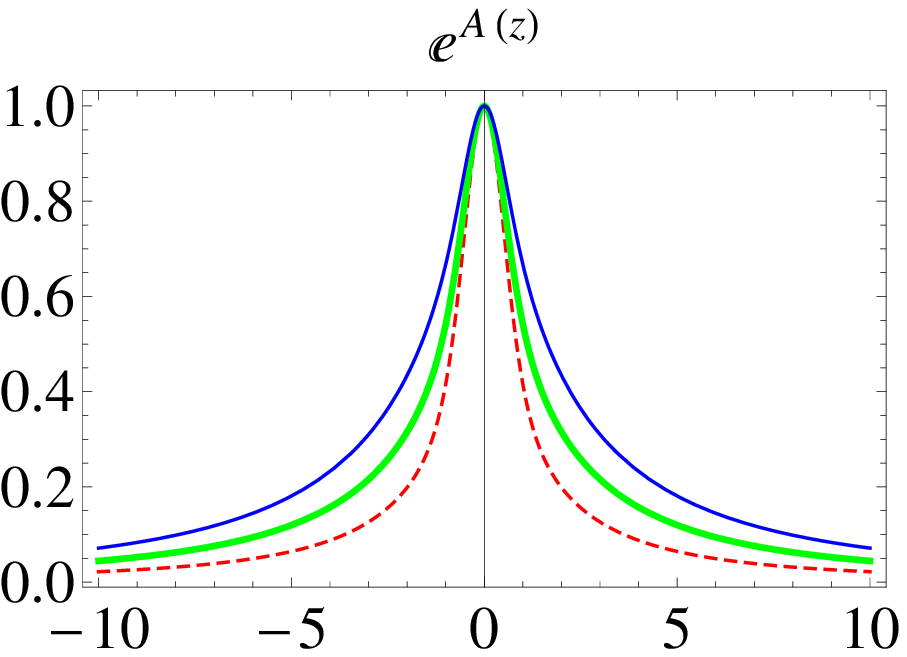}
\includegraphics[scale=0.45]{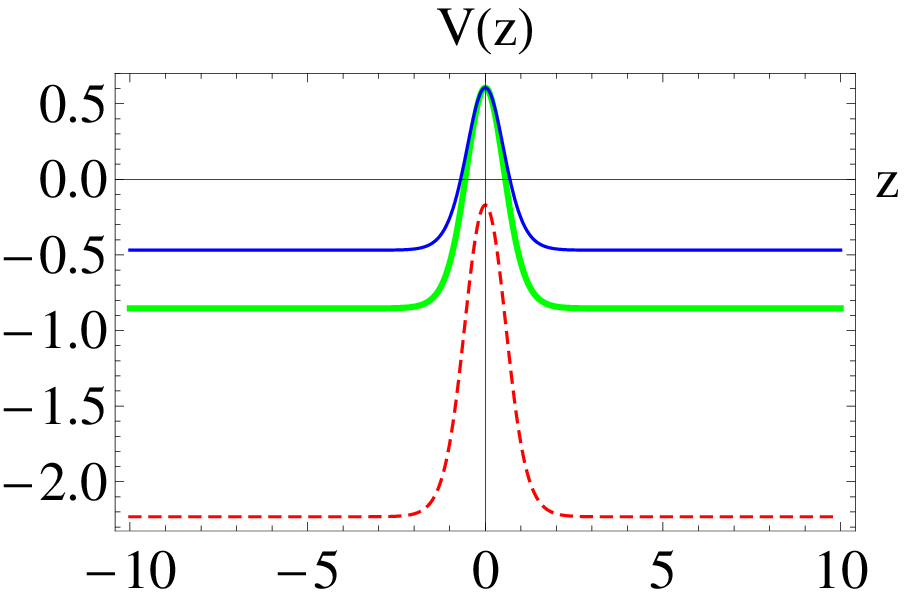}
\includegraphics[scale=0.45]{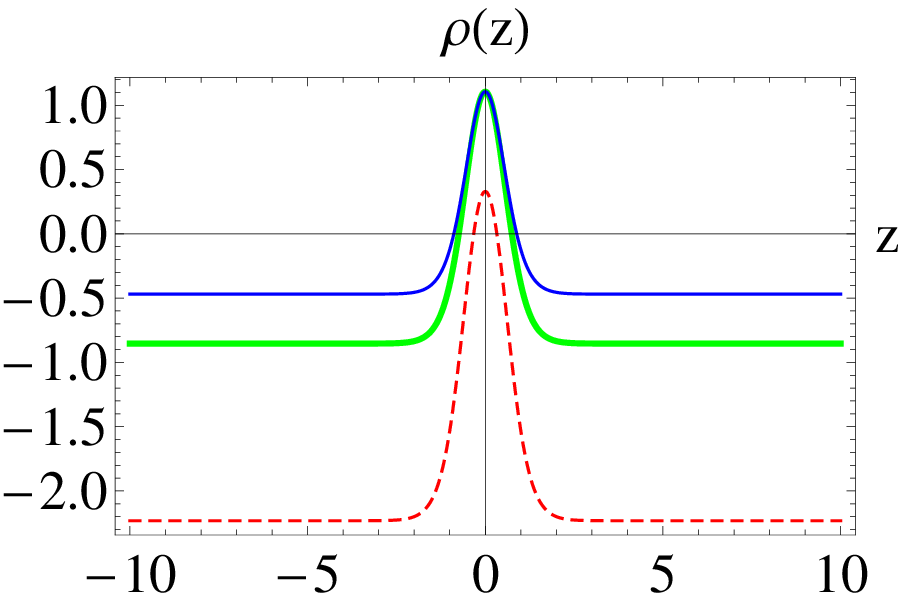}
\includegraphics[scale=0.45]{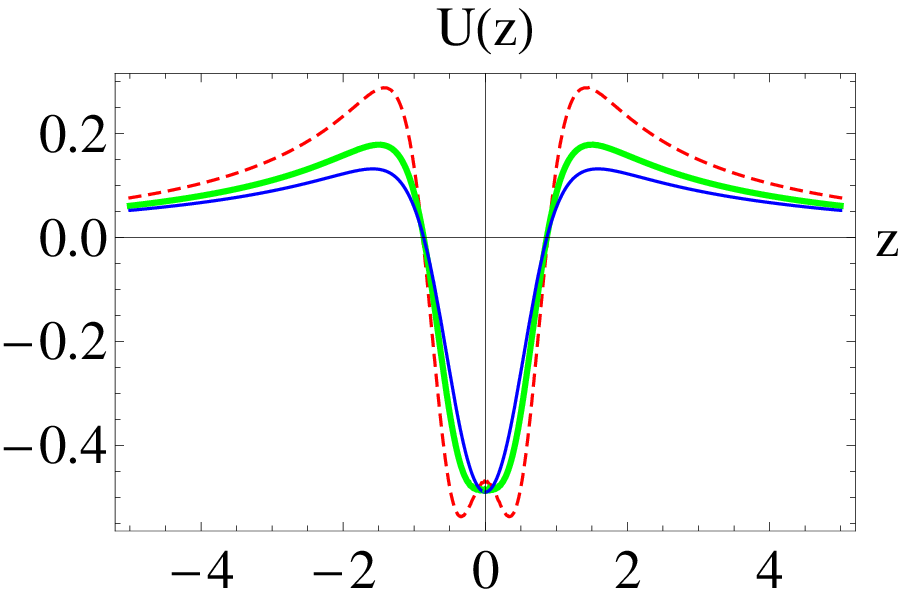}
\caption{The shapes of $A(z)$, $V(\phi(z))$, $\rho(z)$, and $U(z)$ corresponding to $A_-(z)$. The parameters are set to $D=5$, $k=1$, $v=1$, $(\beta_1,\beta_4)=(0.1,-0.05)$ (dashing red lines), $(0.08,-0.05)$ (thick green lines) and $(0.05,-0.05)$ (thin blue lines). \label{picture 2} }
\end{figure}

\begin{figure}[htbp]
\subfigure[$A''_+(0)$]{\label{range 1}
\includegraphics[scale=0.44]{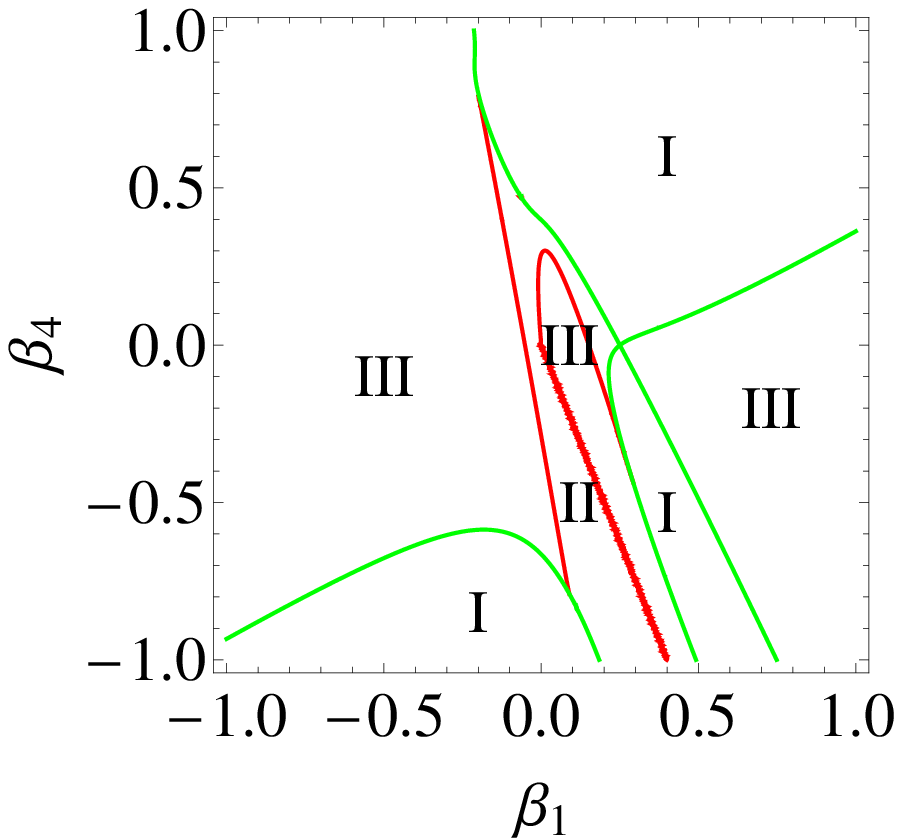}}
\subfigure[$A''_-(0)$]{\label{range 2}
\includegraphics[scale=0.44]{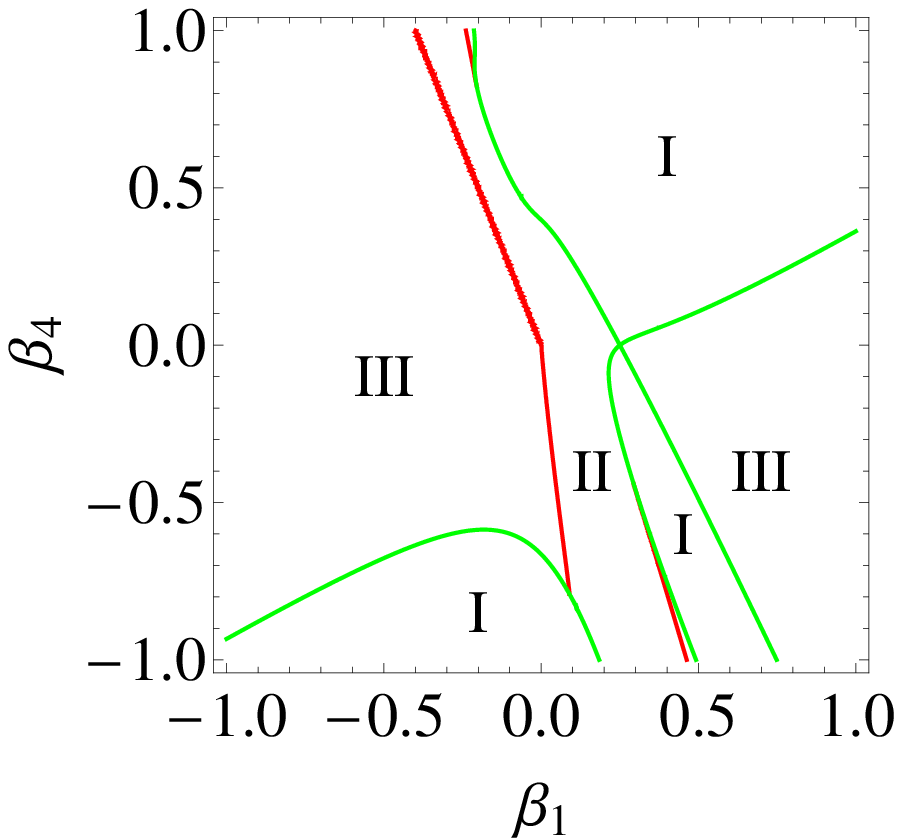}}
\caption{The dependence of $A''_+(0)$ and $A''_-(0)$ on $(\beta_1,\beta_4)$.
Region I: there is no solution.
Region II: $A''(0)<0$.
Region III: $A''(0)>0$.
The parameters are set to $D=5$, $k=1$ and $v=1$.}\label{range}
\end{figure}

It is know that there is only one brane solution in general relativity for a given scalar potential or scalar field. For our brane model with auxiliary fields, there are two brane solutions for a given scalar field.
This can be seen clearly by writing the algebra equation of $A''(0)$ from Eq.~(\ref{WarpFactorEOM}):
\begin{eqnarray}
c_2 A''(0)^{2} + c_1 A''(0) + c_0 = 0. \label{Eq_ddA0}
\end{eqnarray}
Here $c_i$ are coefficients that consist of $k$, $v$, $\beta_1$, and $\beta_4$, and it is not necessary to list their complex expressions here. We set $k=v=1$ in the numerical calculation.
For most sets of $(\beta_1,\beta_4)$, Eq.~(\ref{Eq_ddA0}) gives two solutions:
\begin{eqnarray}
 A''_{\pm}(0)=\frac{-c_1\pm\sqrt{c_1^{2}-4c_2 c_0}}{2c_2}.
\end{eqnarray}
We denote their corresponding warp factors as  $A_+(z)$ (Fig.~\ref{picture 1}) and $A_-(z)$ (Fig.~\ref{picture 2}), respectively.
Figure~\ref{range} shows the dependence of $A''_+(0)$ and $A''_-(0)$ on $(\beta_1,\beta_4)$,
where the meaning of each region is listed as follows:

Region I: $c_1^{2}-4c_2c_0<0$, there is no solution.

Region II: $A''(0)<0$.

Region III: $A''(0)>0$.

\begin{figure}[htbp]
\includegraphics[scale=0.41]{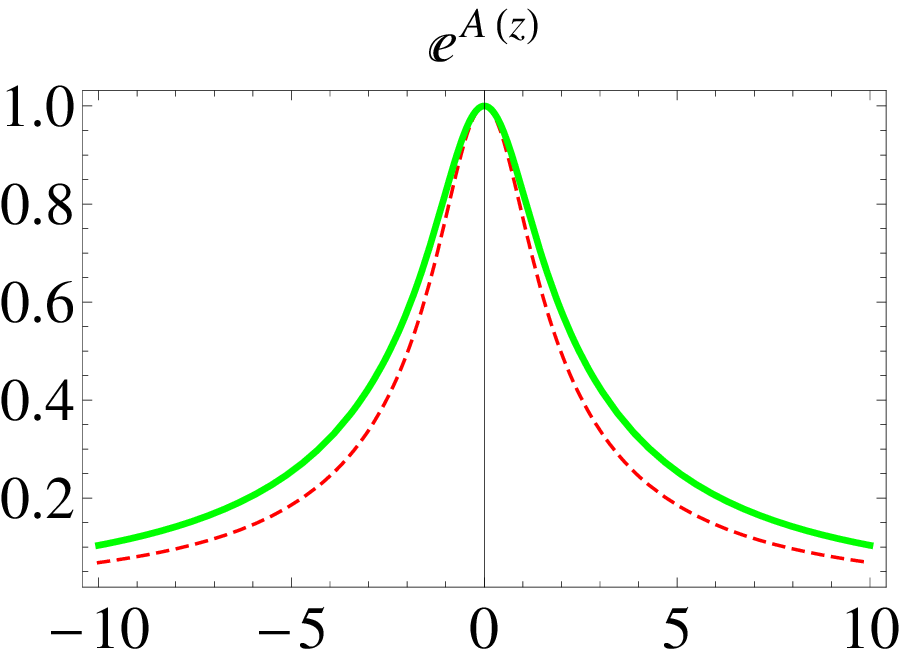}
\includegraphics[scale=0.45]{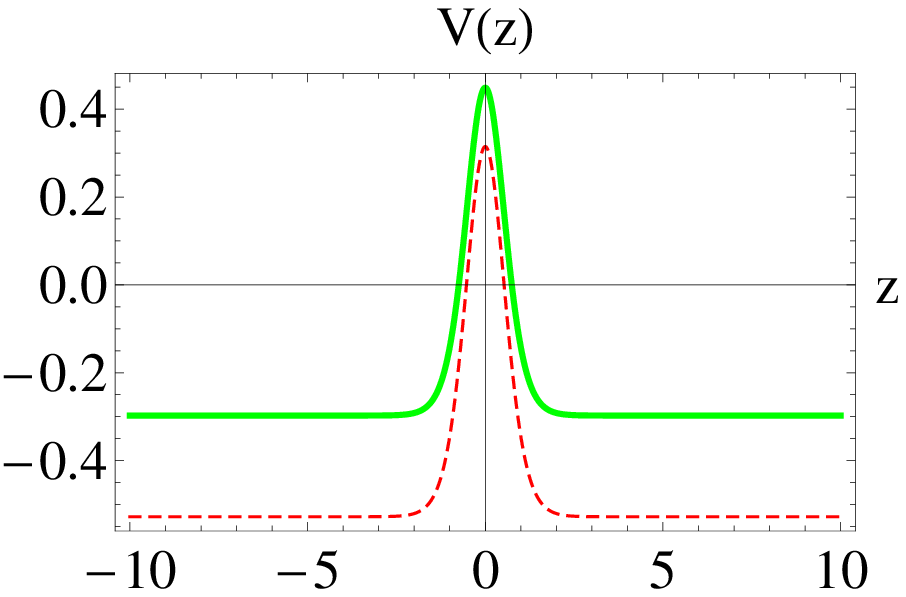}
\includegraphics[scale=0.45]{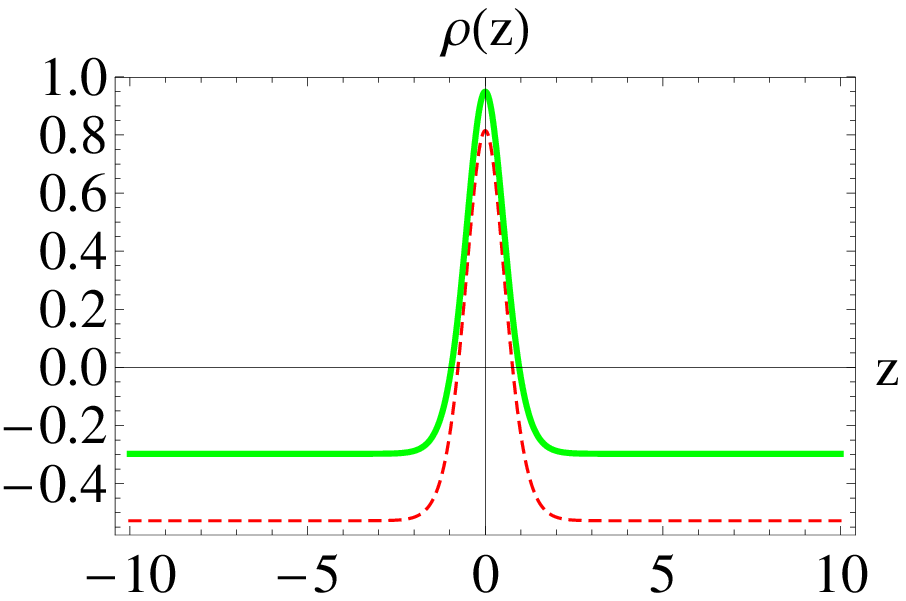}
\includegraphics[scale=0.45]{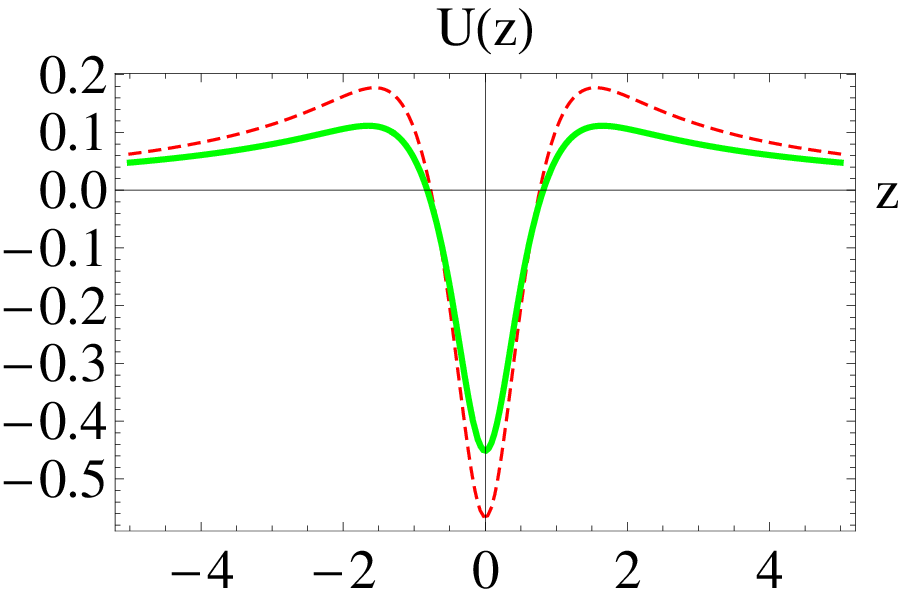}
\caption{The two solutions of $A(z)$, $V(\phi(z))$, $\rho(z)$, and $U(z)$ for a set of parameters $(\beta_1,\beta_4)=(0.03,-0.2)$. Dashing red lines correspond to $A_+(z)$, and thick green lines correspond to $A_-(z)$.
The other parameters are set to $D=5$, $k=1$ and $v=1$.  \label{two physical picture}}
\end{figure}

The warp factor related to region II is an ordinary solution with $A''(0)<0$. However, the warp factor related to region III is a deformed solution with $A''(0)>0$.
We cannot identify whether the solutions corresponding to the points in regions II and III are physical until we numerically calculate their asymptotical behaviors. The physical solutions are those satisfying
$e^{2A(|z|\rightarrow\infty)}\rightarrow0$ because they will guarantee the localization of the four-dimensional massless graviton (the gravity zero mode). In some regions we get two physical solutions as shown in Fig.~\ref{two physical picture}. This is very different from brane models in general relativity.

It can be seen from Fig.~\ref{picture 1} that even though we choose a kink solution $\phi=v\tanh(kz)$, the warp factor $A(z)$ may have an interesting behavior with $A''(0)\geq0$ as in the double kink solution in some other modified gravity theories  \cite{Liu2012,Guo2012}. This deformed warp factor may give some interesting phenomena like resonant KK modes of gravity or scalar field.

Finally, we can get the scalar potential $V(\phi(z))$ in the $z$ coordinate though Eq.~({\ref{V}}) and  the energy density for a static observer $U^A=(e^{-A(z)},0,0,0,0)$:
\begin{eqnarray}\label{EnergyDensity}
\rho(z)=T_{AB}U^A U^B=\frac{1}{2}e^{-2 A}\phi'^{2}(z)+V(\phi(z)).
\end{eqnarray}
Figures~\ref{picture 1}, \ref{picture 2}, and~\ref{two physical picture} show numerical results of
the warp factor $A(z)$, scalar potential $V(\phi(z))$, energy density $\rho(z)$,
and Schr\"odinger-like potential $U(z)$ of the gravitational KK modes for some sets
of $(\beta_1,\beta_4)$.
As shown in these figures, even though the deformation of the warp factor does not effect the energy density (no brane splitting phenomenon), it indeed results in the deformation of the effective potential of the gravitational KK modes. So we may get some new effects on the localization of the matters or gravity \cite{Liu:2009ve} in these cases.

In next section, we mainly focus on the stability of the gravitational perturbation and localization of gravity zero mode, which are two important issues for a brane model that should be investigated before considering their application.

\section{Tensor perturbation and localization of gravity zero mode}\label{tensor perturbation and the localization of gravity}

Now, we consider the tensor perturbation of the background metric, which relates to the spin-2 graviton. The perturbed metric takes as the form
\begin{eqnarray}\label{perturbed metric}
ds^2=a^2(z)\left[(\eta_{\mu\nu}+h_{\mu\nu})dx^{\mu}dx^{\nu}+dz^2\right],
\end{eqnarray}
where the tensor perturbation $h_{\mu\nu}(x,z)$ is transverse and traceless (TT):
\begin{eqnarray}\label{tt condition}
\partial^{\mu}h_{\mu\nu}=0=h^{\lambda}_{~\lambda}.
\end{eqnarray}
The indices are raised and lowered by $\eta^{\mu\nu}$ and $\eta_{\mu\nu}$, respectively.

For the later convenience, we will discuss some general aspects of the tensor perturbation in flat braneworld model for a general modified gravity. For the metric (\ref{metric}), the $\mu\nu$ components of the modified Einstein equations give
\begin{eqnarray}\label{abstract background}
\eta_{\mu\nu}\Theta[A,\phi]=0.
\end{eqnarray}
For our case, $\Theta[A,\phi]$ is given in Eq. (\ref{implicit eq 1}), i.e., $\Theta[A,\phi]=u V^{2}(\phi)+fV(\phi)+g$.
For $h_{\mu\nu}$ is TT, its linear perturbation  equation must take as
\begin{eqnarray}\label{general form}
\cdots+E\partial_{z}\partial_{z}h_{\mu\nu}+B\partial_{z}h_{\mu\nu}+Ch_{\mu\nu}=-\partial^{\lambda}\partial_{\lambda}h_{\mu\nu},
\end{eqnarray}
where the ellipsis denotes the terms involving higher derivatives of the tensor perturbation, and all the coefficients $E$, $B$, $C$ are functions of $z$. In the following, we will show that $C$ will vanish on account of the background modified Einstein equations (\ref{abstract background}). There are some terms that have no contribution to the background equations, but they contribute to the linear perturbation equations. For example, the terms
$\partial_{z}\eta_{\mu\nu}$, $\partial_{z}\partial_{z}\eta_{\mu\nu}$, $\partial_{\lambda}\eta_{\mu\nu}$
in the background equations will give $\partial_{z}h_{\mu\nu}$, $\partial_{z}\partial_{z}h_{\mu\nu}$, $\partial_{\lambda}h_{\mu\nu}$
in linear perturbation equations, respectively.
%
But all their contributions only involve in $E$, $B$, $\partial^{\lambda}\partial_{\lambda}h_{\mu\nu}$, and the higher-derivative terms. So finally, the only term that contributes to $Ch_{\mu\nu}$ is $\Theta[A,\phi] h_{\mu\nu}$, i.e., $C=\Theta[A,\phi]=0$. It should be pointed out that this result depends on the specific form of the metric (\ref{metric}) and the TT conditions of the tensor perturbation (\ref{tt condition}).

Here, we restrict our interest on the theory involving derivatives of
the tensor perturbation only up to second order.
(For some higher-derivative gravity theories such as $f(R)$ and EiBI gravities,
the tensor perturbation equation in the braneworld model may be second order \cite{Liu2011a,Liu2011,Liu2012a}.)
Thus, the right hand side of Eq. (\ref{general form}) is the unique term involving four-dimensional derivatives. Further, with the expansion $h_{\mu\nu}(x,z)=\epsilon_{\mu\nu}e^{-ipx}\Phi(z)$ and $p^{2}=-m^{2}$, Eq. (\ref{general form}) gives
\begin{eqnarray}
E\partial_{z}\partial_{z}\Phi+B\partial_{z}\Phi=-m^{2}\Phi.
\end{eqnarray}
With a coordinate transformation
\begin{eqnarray}
{dw}=\frac{{dz}}{\sqrt{E}} \label{wz}
\end{eqnarray}
and a definition
\begin{eqnarray}
Q=-\frac{1}{2}\frac{\partial_{w}E}{E} + \frac{B}{\sqrt{E}}, \label{Q}
\end{eqnarray}
we have
\begin{eqnarray}
-\partial_{w}\partial_{w}\Phi-Q\partial_{w}\Phi=m^{2}\Phi.
\end{eqnarray}
Then by defining $\Phi=G\tilde\Phi$, where $\partial_{w}G=-\frac{1}{2}QG$, we finally arrive at a Schr\"{o}dinger-like equation
\begin{eqnarray}\label{Sch_eq}
-\partial_{w}\partial_{w}\tilde\Phi
  +\left(\frac{1}{2}\partial_{w}Q+\frac{1}{4}Q^{2}\right)\tilde\Phi=m^{2}\tilde\Phi.
\end{eqnarray}
It can be written as
$\left(\partial_{w}+\frac{Q}{2}\right)
\left(-\partial_{w}+\frac{Q}{2}\right)\tilde\Phi=m^{2}\tilde\Phi,
$
which implies that the eigenvalues are nonnegative, i.e., $m^{2}\geq0$.
Thus, there is no tachyon state, and the brane is stable under the tensor perturbation.

For the zero mode with $m^{2}=0$, the Schr\"{o}dinger-like equation reduces to $\partial_{w}\tilde{\Phi}_0=\frac{Q}{2}\tilde{\Phi}_0$ and the solution is
\begin{eqnarray}\label{ZeroMode}
\tilde\Phi_0 {\propto} e^{\frac{1}{2}{\int}Qdw}.
\end{eqnarray}

For our model, the functions $E$ and $B$ are turned out to be
\begin{eqnarray}
 E &\!=\!& 1+2\beta_4 e^{-2A}\phi'^2, \\
 B&\!=\!& (D-2) A'
   + 2\Big\{
            \beta _4  \left[(D-4) \phi '-2\right] A'\nonumber \\
     &&
           -\beta _1 \left[(D+1) A' \phi '+\phi ''\right]
      \Big\}  e^{-2A}\phi' .
\end{eqnarray}

Now we investigate the localization of the gravitational zero mode (\ref{ZeroMode}).
We only need to analyze its asymptotic behavior at the boundary.
With $\partial_{z}\phi\rightarrow 0$ and $V(\phi)\rightarrow V_{0}$ as $z\rightarrow \infty$, we have
\begin{eqnarray}
\tilde\Phi_0(z\rightarrow \infty)\rightarrow e^{(D-2)A/2}.
\end{eqnarray}
Then, by a coordinate transformation $\frac{dy}{dz}=e^{A}$ from the conformal coordinate $z$ to the physical coordinate $y$, it can be shown that the asymptotic solution of $A$ is $A\rightarrow -\sqrt{-\frac{1}{6}\Lambda_{\text{eff}}}~|y|$ with $\Lambda_{\text{eff}}$ the effective cosmological constant. So the integral $\int \tilde\Phi_0^2 dw \rightarrow\int e^{(D-3)A} dy$ converges if $D>3$. So the gravitational zero mode can be localized on the brane, and hence, this ensures that the four-dimensional Newtonian potential can be realized on the brane in the low energy limit.

From the Schr\"{o}dinger-like equation (\ref{Sch_eq}), we get the effective potential for the gravitational KK modes:
\begin{eqnarray}
U=\frac{1}{2}\partial_{w}Q+\frac{1}{4}Q^{2}.\label{Uw}
\end{eqnarray}
In the conformal coordinate $z$, we have from (\ref{Q}) and (\ref{wz}):
\begin{eqnarray}
Q(z)=\frac{2 B-{E'}}{2 \sqrt{E}}.
\end{eqnarray}
Then the effective potential (\ref{Uw}) can be reexpressed in the conformal coordinate as
\begin{eqnarray}
U(z)=\frac{2 B'-E''}{4}
  +\frac{4 B^2-8 B E'+3 E'^2}{16 E},
\end{eqnarray}
from which it is easy to show that $U(z)\rightarrow 0$ as $z\rightarrow\infty$. The numerical results for the effective potential are shown in Figs. \ref{picture 1}, \ref{picture 2}, and \ref{two physical picture}, from which it can seen that the potential has a shape of volcano and vanishes at the boundary of the extra dimension. Thus, besides the massless bound state, there are continuous modes, which are nonlocalized massive gravitons. The four-dimensional gravity potential on the brane is determined by the interaction between these gravitons and the matter on the brane \cite{Csaki2000,Rubakov2001,Arkani-Hamed2000a,Csaki2004,Bazeia2009}. The massless graviton generates the Newtonian potential, while the massive gravitons result in the correction.
The solutions of the massive KK modes depend on the shape of the effective potential $U(z)$.
From Fig.~\ref{picture 1}, we can see that the effective potential $U(z)$ may has one (thick green line) or three (dashing red line) potential wells when the warp factor is deformed.
From Fig.~\ref{picture 2}, it can be seen that $U(z)$ may has one (thick green and thin blue lines) or two (dashing red line) potential wells when the warp factor is nondeformed.
In these cases, even though the energy density of the brane does not split, there are many interesting results related to the deformation of the effective potential $U(z)$.

\section{Conclusion}\label{Conclusion}

In this paper, we investigated the thick braneworld model in gravity with auxiliary fields.
By numerically studying the model with a kink configuration of the scalar field, 
we found that the solutions of the brane system with nondeformed and deformed warp factors even though the energy density of the brane does not split.
This new phenomenon is connected to the presence of auxiliary fields.
We also gave the parameter spaces corresponding to these two types of warp factors.

The tensor perturbation of the background metric of the flat brane system was analyzed. 
Although the gravity with auxiliary fields we considered in this paper is a fourth-order derivative theory, the linear equation of the tensor perturbation is of second order. 
The equation of motion of the KK mode of the tensor perturbation was turned into a Schr\"odinger-like equation. The effective potential $U(z)$ has the shape of a volcano with one or more potential wells, which do not depend on whether or not the warp factor is deformed.
It was shown that the tensor perturbation of the flat brane model is stable.
The massless mode of the tensor perturbation is localized on the brane,
while the massive modes are continuous and nonlocalized.
Therefore, the four-dimensional Newton's gravity can be realized on the brane.


\begin{acknowledgements}
We thank Prof. Y.Q. Wang and Drs. B.M. Gu, X.L. Du, and F.W. Chen  for helpful discussions.
This work was supported by the National Natural Science Foundation of China (Grant No. 11375075). K. Yang was supported by the Scholarship Award for Excellent Doctoral Student granted by Ministry of Education.
\end{acknowledgements}

\end{document}